\documentclass{elsart}
\usepackage{graphicx}
\usepackage{epstopdf}
\begin{document}
\begin{frontmatter}
\title{Recursive renormalization of the singlet
one-pion-exchange plus point-like interactions}
\author[CESET]{V. S. Tim\'oteo}
\author[ITA]{T. Frederico,}
\author[UFF]{A. Delfino,}
\author[IFT]{Lauro Tomio}
\address[CESET]{Centro Superior de Educa\c c\~ao Tecnol\'ogica,
Universidade Estadual de Campinas,
13484-370, Limeira, SP, Brasil}
\address[ITA]
{Departamento de F\'\i sica, Instituto Tecnol\'ogico de
Aeron\'autica, CTA, 12228-900, S\~ao Jos\'e dos Campos, Brasil}
\address[UFF]
{Instituto de F\'\i sica, Universidade Federal Fluminense,
24210-900 Niter\'oi, RJ, Brasil}
\address[IFT]{Instituto de F\'\i sica Te\'orica, Universidade
Estadual Paulista,  01405-900, S\~{a}o Paulo, Brasil }
\date{\today}
\maketitle
\begin{abstract}
The subtracted kernel approach is shown to be a powerful
method to be implemented recursively in scattering equations with
regular plus point-like interactions. The advantages of the method
allows one to recursively renormalize the potentials, with higher
derivatives of the Dirac-delta, improving previous results. The
applicability of the method is verified in the calculation of the
$^1S_0$ nucleon-nucleon phase-shifts, when considering a potential
with one-pion-exchange plus a contact interaction and its
derivatives. The $^1S_0$ renormalization parameters are fitted to
the data. The method can in principle be extended to any
derivative order of the contact interaction, to higher partial
waves and to coupled channels.
\end{abstract}
\end{frontmatter}

\section{Introduction}

The pioneering work of Weinberg~\cite{wei} launched the basis for
the effective field theory (EFT) of nuclear forces starting from
the expansion of an effective chiral Lagrangian. It gives a
nucleon-nucleon (NN) interaction which is, in leading order, the
one-pion-exchange potential (OPEP) plus a Dirac-delta. The program
of applying effective field methods to the NN system was pursued
by many authors with significative results in few-nucleon systems
(see e.g.~\cite{bira}). In a more general context of few-body
systems, short range interactions have many applications which are
discussed in detail in ref.~\cite{fedorov}.

More recently, the authors of \cite{nieves} and \cite{arriola}
treat the one-pion-exchange potential (OPEP) plus derivative
Dirac-delta interactions using dimensional and boundary condition
regularizations, respectively. In particular, when comparing the
theoretical approach with the data~\cite{nij}, it was shown in
ref.~\cite{fred99} that the leading order interaction, OPEP plus a
Dirac-delta,  renormalized using subtracted scattering equations,
dominates the results obtained for the $^3S_1-^3D_1$ phase-shifts
and mixing parameter. However, considering only the leading-order
term, the results obtained for the $^1S_0$ phase shift are not
satisfactory. This is an evidence that higher order terms in the
effective interaction are important in this channel (see e.g.
refs.~\cite{nieves,arriola}).

The fit of the $^1S_0$ phase-shift up to laboratory momentum
$p_{Lab}\sim 300$~MeV/c requires an effective NN interaction with
the addition of a term with second order derivatives of the
Dirac-delta. In the relative momentum space, we have
\begin{eqnarray}
\langle\vec {p'}|V|\vec p \rangle = \langle\vec {p'}|V^s_\pi|\vec
p \rangle &+& \sum_{i,j=0}^1\lambda_{ij}{p'}^{2i}{p}^{2j},
\label{veft}
\end{eqnarray}
where the $\lambda$'s are unregulated strengths and $\langle\vec
{p'}|V^s_\pi|\vec p \rangle$ is a matrix element of the
one-pion-exchange potential. The motivation of the second term of
(\ref{veft}) is to simulate effects of heavy particle exchanges by
a sum of a Dirac-delta interaction and its derivatives, while
keeping the OPE as the long range part of the interaction.
  In the renormalization method described in \cite{fred99},  we
include $\lambda_{11}$, as the method is based on a
kernel-subtraction procedure that generates terms of the type
$p'^2p^2$ in the scattering matrix elements even when one
considers only $p^2$ and $p'^2$ in the interaction. In fact, the
method also generates higher-order derivatives terms when OPE is
considered. In the Weinberg counting rule scheme, the derivative
contact interactions we are considering comes along with the
two-pion-exchange (TPE)~\cite{wei,ordonez}. For a recent
calculation in the nn system considering TPE see e.g.
ref.~\cite{epelbaum}. As an exploratory calculation, in the
present work we consider only the contact terms, leaving TPE for a
future work.

In the scattering equation, the effective bare
potential~(\ref{veft}) produces integrals that diverge as much as
$p^5$. Therefore it is necessary at least three subtractions in
the kernel of the Lippman-Schwinger (LS) equation, since each
subtraction introduces a factor of $p^{-2}$. Differently from the
recent works~\cite{nieves} and \cite{arriola}, we implement the
method of subtracted scattering equations~\cite{plb00} to deal
with this problem. The subtraction method has also been shown to
be practical in providing renormalization group invariant
solutions for three-body scattering equations with contact
interactions~\cite{adh95,recomb,afnan}, and also proved to be
useful in describing the halo structure of weakly bound exotic
light nuclei~\cite{halo}.

The one-subtracted scattering equation used in our previous
work~\cite{fred99} was generalized in~\cite{plb00} to allow any
order of subtraction, permitting the inclusion of derivatives of
the contact interaction in the effective two-body potential. The
driving term of the n-subtracted LS equation is constructed
recursively, renormalizing the model at each subtraction order,
while keeping renormalization group invariance of the approach
(for details see~\cite{plb00}).

In the present work, we obtain the $^1S_0$ nucleon-nucleon
amplitude from the effective interaction (\ref{veft}), using a
three-times subtracted  scattering equation, as demanded by the
higher divergent term of the type $p^5$. We perform an analysis of
the physical contribution coming from each order term in the
recursive order-by-order renormalization procedure. It this way we
access their significance in the parameterization of the effective
interaction to obtain the desired observables.

We should note that the contact interactions that we are
considering are obviously meaningless without regularization and
renormalization  due to the generated ultraviolet divergences in
the Lippman-Schwinger equation.

This work is organized as follows. In section 2, we discuss the
subtraction method applied to OPEP.  In section 3, we present the
main formulas of the recursive subtraction method to treat the
scattering equation, as well as our strategy to solve it. In section
4, we show numerical results for the singlet phase-shift with
our conclusion.

\section{One-fold subtracted T-matrix equation: the one-pion-exchange potential}

The partial wave decomposition of OPEP for the $^1S_0$ state
is~\cite{BrownJack}:
\begin{eqnarray}
V_{\pi , s}(p^\prime,p) &=& \frac{g_a^2}{16\pi f_\pi^2}
-\frac{g_a^2}{32\pi f_\pi^2} \int^1_{-1}dx
\frac{m_\pi^2}{p^2+{p'}^2-2 p {p'}x+m_\pi^2}  ~, \label{sing}
\end{eqnarray}
where $p$ and $p^\prime$ are the relative momentum. The regular
part of OPE has a finite scattering matrix, solution of the
partial-wave projected LS equation:
\begin{eqnarray}
T_\pi(p',p;k^2)= V_{\pi,s}^{reg}(p',p) +
\frac{2}{\pi}\int^\infty_0dqq^2\frac{V_{\pi,s}^{reg}(p',q)}{k^2-q^2+i\epsilon}
T_\pi(q,p;k^2)~,\label{tpifinal}
\end{eqnarray}
where  $V^{reg}_{\pi ,s}(p^\prime,p)$ is the second term in the
right-hand side of (\ref{sing}).

The T-matrix of the one-pion-exchange plus the Dirac-delta
potential~\cite{fred99} is obtained here using Distorted Wave
Theory~\cite{dwt} as suggested in \cite{nieves}:
\begin{eqnarray}
T_{\pi + \delta}(E)= T_{\pi}(E)+ \left[
1+T_{\pi}(E)G^{(+)}_0(E)\right] T_\delta(E)
\left[1+G^{(+)}_0(E)T_{\pi}(E) \right] \ , \label{TPIVW1}
\end{eqnarray}
where $G^{(+)}_0(E)=[E-H_0+i\epsilon]^{-1}$ is the free Green's
function with $H_0=p^2$ in the two-nucleon rest-frame, and
\begin{eqnarray}
T_\delta(E)= V_\delta+ V_\delta G^{(+)}_{\pi}(E)T_\delta(E)~.
\label{TPIVW2}
\end{eqnarray}
The Green's function for the regular part of the pion-exchange
interaction is
\begin{eqnarray}
G^{(+)}_{\pi}(E)=G^{(+)}_0(E)+G_0^{(+)}(E)T_{\pi}(E)G^{(+)}_0(E)~.
\end{eqnarray}

The direct solution of (\ref{TPIVW2}) is plagued by infinities
in the momentum integration. In order to get finite solution
within our scheme we reformulate (\ref{TPIVW2}) in a
subtracted fashion, where only one subtraction is
enough~\cite{fred99}.

Let us briefly explain the subtraction
method~\cite{fred99,plb00} discussing the treatment of potentials
where the matrix elements behave as a constant in the ultraviolet
momentum region, like in (\ref{TPIVW2}), which needs one
subtraction to allow finite results for the T-matrix.

Using the Lippmann-Schwinger equation
\begin{eqnarray}
T(E)= V[1+G^{(+)}_0(E)T(E)] = [1+T(E)G^{(+)}_0(E)]V  \
,\label{RTV1}
\end{eqnarray}
the potential $V$ can be formally written in terms of the T-matrix
at a certain energy $-\mu^2$ as
\begin{eqnarray}
V= \left[{1+T(-\mu^2) G_0(-\mu^2)}\right]^{-1}T(-\mu^2) \ .
\label{RV}
\end{eqnarray}
For convenience we use a negative energy for the subtraction such
that the Green's function is real.

>From (\ref{RV}) and (\ref{RTV1}) we obtain the one-subtracted
scattering equation:
\begin{eqnarray}
T(E)&=&T(-\mu^2)+T(-\mu^2) \left( G^{(+)}_0(E)-
G_0(-\mu^2)\right)T(E)\ . \label{RT1V0} \end{eqnarray}

By introducing a convenient notation, which is useful for more
subtractions, the driving term of (\ref{RT1V0}) is denoted as
$V^{(1)}(-\mu^2)\equiv T(-\mu^2)$. A subtracted free Green
function is written as
\begin{equation}
G^{(+)}_{1}(E;-\mu^2) \equiv G^{(+)}_0(E)- G_0(-\mu^2) =-(\mu^2+E)
G_0(-\mu^2) G^{(+)}_0(E), \label{G1} \end{equation} which allows
to rewrite the renormalized T-matrix equation~(\ref{RT1V0})  as
\begin{eqnarray}
T(E)&=& V^{(1)}(-\mu^2)+ V^{(1)}(-\mu^2)G^{(+)}_{1}(E;-\mu^2)T(E)
\label{RT10} . \end{eqnarray} For a Dirac-delta potential,
(\ref{RT10}) produces finite results once $V^{(1)}(-\mu^2)$ is
given.  In this simple case, the matrix elements of
$V^{(1)}(-\mu^2)$ in the relative momentum representation is just
the renormalized coupling constant.

The method of subtracted equations  is easily applied to (\ref{TPIVW2}),
and one should note that the free Green's function in (\ref{RT10}) is
being replaced by the interacting one, $G_\pi$.
We obtain the T-matrix for an arbitrary energy $E$ in terms of the
T-matrix at a given energy scale $-\mu^2$:
\begin{eqnarray}
T_\delta(E)&=&T_\delta(-\mu^2)+T_\delta(-\mu^2) \left(
G^{(+)}_\pi(E)- G_\pi(-\mu^2)\right)T_\delta(E)~. \label{T1V0}
\end{eqnarray}

The renormalized strength of the interaction defines $T_\delta(E)$
at the subtraction point, $T_\delta(-\mu^2)=\lambda_{\mathcal{ R}00}$,
which is enough to get a finite amplitude for the OPEP plus a
Dirac-delta at this energy:
\begin{eqnarray}
T_{\pi + \delta}(p',p&;&-\mu^2)= T_{\pi}(p',p\,;-\mu^2) \nonumber \\
&+& \left[ 1+\frac{2}{\pi}\int^\infty_0dq~q^2
\frac{T_{\pi}(p',q;-\mu^2)}{-\mu^2-q^2}\right]\lambda_{\mathcal{ R}00}
\left[1+\frac{2}{\pi}\int^\infty_0dq'~{q'}^2
\frac{T_{\pi}(q',p;-\mu^2)}{-\mu^2-{q'}^2} \right] \ ,
\label{TPIVW3}
\end{eqnarray}
where the T-matrix for the regular part of the one-pion-exchange
potential in the $^1S_0$ channel is given by eq.~(\ref{tpifinal}).

\section{Subtracted T-matrix equations}

The scattering T-matrix is not finite for the bare effective
potential (\ref{veft}), which imply that regularization and
renormalization are required to define the scattering amplitude.
The method we use consists in constructing regularized and
renormalized scattering equations with propagators subtracted at
certain scales, which are convenient for introducing the physical
inputs.

Our motivation in using the subtracted method was to treat a bare
potential that includes a regular part plus the Dirac-delta
interaction and its derivatives, for which the T-matrix has a
definite form for a certain $E=-\mu^2$. That given, no
approximations are performed in getting the scattering amplitude.

 Let us consider point-like interactions that include up to four
terms, a contact interaction plus its derivatives. After
partial-wave decomposition to the singlet $s-$wave state, the bare
potential is given by
\begin{eqnarray}
V_{s}(p^\prime,p) &=& V^{reg}_{\pi ,s}(p^\prime,p)
+\sum_{i,j=0}^{1}\lambda_{ij} {p'}^{2i} p^{2j}
\;\;\;(\lambda_{ij}=\lambda_{ji}^*) \nonumber\\
&=& \underbrace{V^{reg}_{\pi
,s}(p^\prime,p)+\lambda_{00}}_{V_{\pi+\delta}} \;+\;
\underbrace{\lambda_{01}{p'}^2+\lambda_{10}p^2+\lambda_{11}{p'}^2p^2}_{V_{\delta'}}
\; , \label{vefts}
\end{eqnarray}
where $V_{\pi+\delta}$ corresponds to the regular part of OPEP plus a
Dirac-delta interaction $V_\delta$.

 In order to calculate the scattering amplitude for the bare
potential (\ref{vefts}) with subtracted equations, the strategy is
the following. First one has to define the starting point of the
iterative procedure to get the driving term of the subtracted
equation. In the present case, at least three subtractions are
required to allow a finite result for the the T-matrix, and two
iterations are consequently needed to get the driving term. The
iterative process begins with the T-matrix of OPE plus Dirac-delta
potentials (\ref{TPIVW3}), leading to one subtraction in the
kernel ($n=1$), as shown in section 2. Next, as we show in the
following subsection, two iterations are performed and added the
amplitude $\lambda_{\mathcal{ R}10}({p'}^2+p^2)
+\lambda_{\mathcal{ R}11} {p'}^2p^2$ at the subtraction scale
($\lambda_{\mathcal{ R}10}$ and $\lambda_{\mathcal{ R}11}$ are the
renormalized strengths of the interaction).

\subsection{Three-fold subtracted T-matrix equation}

A three-fold subtracted T-matrix equation has to be used to define
the scattering amplitude of the effective potential
(\ref{vefts}), which in operator form and for a general number
of subtractions $n$ reads~\cite{plb00}:
\begin{eqnarray}
T(E)&=& V^{(n)}(-\mu^2;E) +
V^{(n)}(-\mu^2;E)G^{(+)}_n(E;-\mu^2)T(E) , \label{tren}
\end{eqnarray}
where
\begin{eqnarray}
V^{(n)}(-\mu^2;E)&\equiv&\left[1-(-\mu^2-E)^{n-1}V^{(n-1)}(-\mu^2;E)
G^{n}_0(-\mu^2)\right]^{-1}V^{(n-1)}(-\mu^2;E)\; , \label{VN} \\
G^{(+)}_n(E;-\mu^2)&\equiv&\left[(-\mu^2-E)G_0(-\mu^2)\right]^n
G^{(+)}_0(E) . \label{GN}
\end{eqnarray}
One should note that the form of the above equation with $n=1$
when applied to three-body systems results in the subtracted form
used in \cite{afnan}, where the boundary condition was taken at
the scattering threshold.

The recursive formula to derive the driving term $V^{(3)}(-\mu^2)$
starts from $V^{(1)}(-\mu^2)=T_{\pi+\delta}(-\mu^2)$  given by
(\ref{TPIVW3}). So far, only the OPEP and the Dirac-delta
interactions have been introduced in the calculation. The higher
order derivatives of the Dirac-delta potential in
(\ref{vefts}) are introduced in the driving term of the
three-fold subtracted equation.

The matrix element of $V^{(3)}_{\pi+\delta+\delta'}(-\mu^2)$ for
the full effective interaction of (\ref{vefts}) in the angular
momentum basis,
\begin{eqnarray}
V^{(3)}_{\pi+\delta+\delta'}(p',p\,;-\mu^2;k^2)=
V^{(3)}_{\pi+\delta}(p',p\,;-\mu^2;k^2) + \lambda_{\mathcal
{R}10}({p'}^2+p^2) +\lambda_{\mathcal{ R}11} {p'}^2p^2~,
\label{vrenfinal}
\end{eqnarray}
get the contribution from the derivatives of the Dirac-delta
interaction, through the values of $\lambda_{\mathcal{ R}ij}$ - the
renormalized strengths of the corresponding   terms in the
potential.

The integral equations for the matrix elements,
$V^{(n)}_{\pi+\delta}(p',p\,;-\mu^2;k^2)$, are
\begin{eqnarray}
V^{(n)}_{\pi+\delta}(p',p&;&-\mu^2;k^2)=
V^{(n-1)}_{\pi+\delta}(p',p\,;-\mu^2;k^2) \nonumber \\
&+& \frac{2}{\pi}\int^\infty_0dqq^2
\left(\frac{\mu^2+k^2}{\mu^2+q^2}\right)^{n-1}
\frac{V^{(n-1)}_{\pi+\delta}(p',q;-\mu^2;k^2)}{-\mu^2-q^2}
V^{(n)}_{\pi+\delta}(q,p;-\mu^2;k^2) , \label{vrenpid}
\end{eqnarray}
with $k=\sqrt{E}$.

The explicit form of the three-fold subtracted LS equation is:
\begin{eqnarray}
T(p',p;k^2)&=&
V^{(3)}_{\pi+\delta+\delta'}(p',p;-\mu^2;k^2) \nonumber \\
&+& \frac{2}{\pi}\int^\infty_0dqq^2
\left(\frac{\mu^2+k^2}{\mu^2+q^2}\right)^3
\frac{V^{(3)}_{\pi+\delta+\delta'}(p',q;-\mu^2;k^2)}{k^2-q^2+i\epsilon}
T(q,p;k^2) \, .
\label{trenfinal}
\end{eqnarray}
In the angular momentum basis the scattering amplitude is
\begin{eqnarray}
T(k,k;k^2)= -\frac{1}{k\cot\delta-i~k}~, \label{sca}
\end{eqnarray}
and the low-energy parameters are defined by the effective range
expansion $ k\cot\delta=-{1}/{a}+(r_0/2)k^2+~. ~. ~. ~,$ with $a$
the scattering length and $r_0$ the effective range.

 We observe that in our method the kernel is subtracted at a
given scale which allows to perform the momentum integrals up to
infinity, in that sense the subtracted equations are automatically
regularized. In the cutoff regularization, the Hilbert space is
truncated and intermediate states goes up to a given momentum
scale, such that the effective interaction includes the effect of
the neglected states~\cite{bogner}. Our method softens the
contribution of the higher momentum intermediate states, with the
partially removed physics parameterized through the driving term.
Of course we could move the subtraction scale, at the expense of a
more complicated form for the driving term. This fact also happens
in a cut-off regularization when the cut-off is moved~\cite{bogner}.

\subsection{Invariance under dislocation of the subtraction point}

The nucleon-nucleon observables are invariant under the change of
the arbitrary subtraction point, therefore one can start at any
convenient energy scale $\mu^2$. However, the form of the driving
term and its coefficients in (\ref{vrenfinal}) which define
the scattering amplitude are tied to the prescription used to
define the renormalized theory. The key point of the
renormalization group method is to change this prescription
without altering the predictions of the theory~\cite{weinbook,justin}.

The invariance of the T-matrix under changes of renormalization
prescriptions imposes a definite rule to modify $V^{(n)}$ in
(\ref{tren}), which appears in a form of a non-relativistic
Callan-Symanzik (NRCS) equation~\cite{plb00,CS}:
\begin{eqnarray}
\frac{\partial V^{(n)}(-{ \mu}^2;E)}{\partial \mu^2} =
-V^{(n)}(-{ \mu}^2;E) \frac{\partial G^{(+)}_n(E;-{ \mu}^2)
}{\partial \mu^2} V^{(n)}(-{ \mu}^2;E) \ , \label{ss3}
\end{eqnarray}
derived from (\ref{tren}) and $ \frac{\partial T(E)}{\partial
\mu^2} =0$. Equation~(\ref{ss3}) substantiate the invariance of
the renormalized T-matrix under dislocation of the subtraction
point, with the boundary condition given  by
(\ref{vrenfinal}). Then, we observe that there is a
non-trivial dependence on the subtraction point appearing in the
driving term of the subtracted scattering equation, although the
physical results of the model are kept unchanged.

The solution of (\ref{ss3}) implies  in a complicated
evolution of $V^{(n)}$ as $\mu$ changes. Not only the lambda's
would change, but also the form of the driving term. The
ultraviolet behavior of the driving term is not changed by the
evolution in $\mu$. The evolution should not be truncated as $\mu$
is varied to keep the T-matrix invariant. At different $\mu$ the
potential $V^{(n)}(-\mu^2;E)$ has a complicate form from the
solution of NRCS equation. Similarly, the evolution of
renormalization group equations as introduced by Bogner, Kuo and
Schwenk~\cite{bogner} for the NN scattering does not truncate on
certain operators as the cutoff is varied  to keep observables
unchanged.

\section{Numerical Results and Conclusion}

The renormalized strengths $\lambda_{\mathcal{ R}00}$,
$\lambda_{\mathcal {R}10}$, $\lambda_{\mathcal{ R}11}$ and the
subtracion point $\mu$ are found by fitting some low-energy phase
shift data. For each set of $\lambda_{\mathcal{ R}10}$,
$\lambda_{\mathcal{ R}11}$ and $\mu$, we fit the singlet
scattering length $a_s=-23.739$~fm through the value of
$\lambda_{\mathcal{ R}00}$. The three parameters left are adjusted
to reproduce the Nijgemen data~\cite{nij} up to the center of mass
momentum $p$ of about 300~MeV/c. We get  an effective range of
$r_{0,s}=2.73$~fm compared with the value of 2.68~fm from
Ref.~\cite{nij}. The parameters found are displayed in
Table~\ref{tab1}, where $\mu = 214$~MeV/c corresponds to a
negative subtraction energy point $\approx$ - 50 MeV.
We emphasize that the
results for the scattering length shown in the table correspond to
the values obtained considering only the parameters found from the
best fit of the singlet phase-shift using Eq.(\ref{vrenfinal}) as
input for the three-fold subtracted scattering equation
(\ref{trenfinal}). Of course, from pionless EFT, we know that one
could adjust the scattering length and the singlet phase-shifts at
very low-energies with only contact interactions.

\begin{table}[t]
\begin{center}
\begin{tabular}{ccccc}
\hline \hline Interaction & $\mu\lambda_{\mathcal{R}00}$ &
$\mu^3\lambda_{\mathcal{R}01}$ & $\mu^5\lambda_{\mathcal{R}11}$ &
$a_s$ [fm] \\ \hline
$\pi$  & 0 & 0 & 0 & -0.878  \\
$\delta$ & -0.1465 & 0 & 0 & -0.135  \\
$\delta^\prime$ & 0 & 4.7124 & 5.0265 & -5.951 \\
$\delta + \delta^\prime$ & -0.1465 & 4.7124 & 5.0265 & -6.995 \\
$\pi + \delta$ & -0.1465 & 0 & 0 & -1.250 \\
$\pi + \delta + \delta^\prime$ & -0.1465 & 4.7124 & 5.0265 & -23.739 \\
\hline \hline
\end{tabular}
\end{center}
\caption{Dimensionless renormalized strenghts and the contributions to
the scattering length.}
\label{tab1}
\end{table}

\begin{figure}[b]
\includegraphics[width=14cm ,height=10cm]{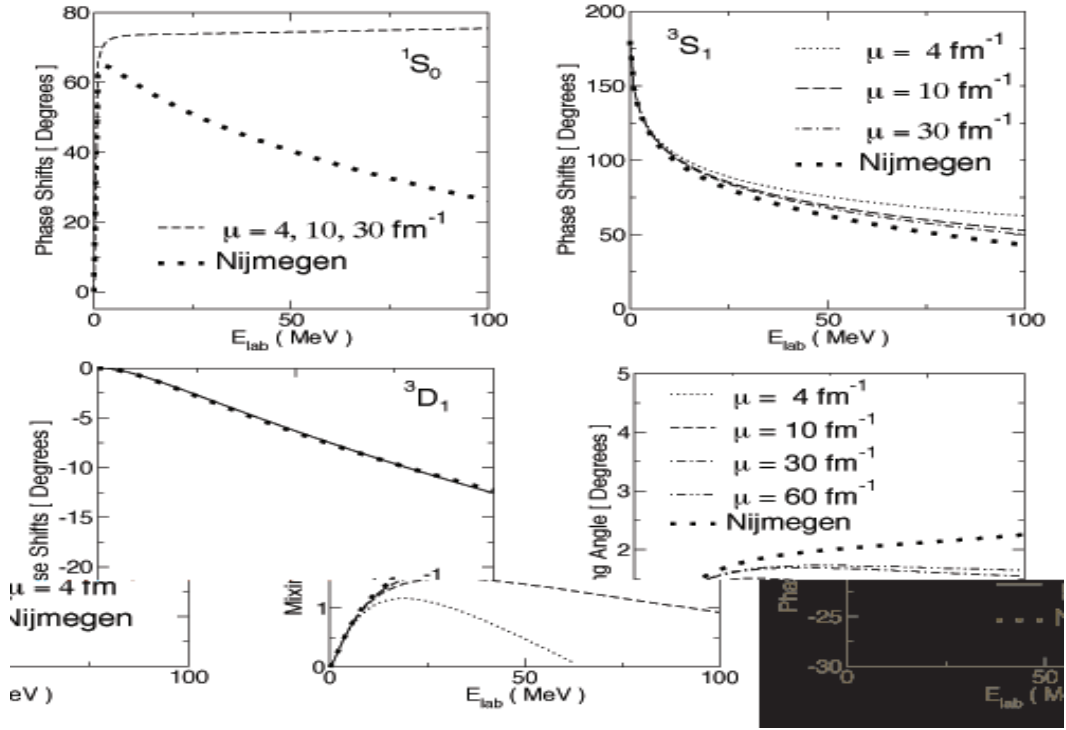}
\caption[dummy0] {Phase shifts for the singlet ($^1S_0$) and the
coupled ($^3S_1 - ^3D_1$) channels for different values of $\mu$
(given in $fm^{-1}$) with only $V_{\pi + \delta}$. The plots were
taken from our previous work \cite{fred99}.} \label{fig1}
\end{figure}

 First, we show in Figure~\ref{fig1} the results of our method
for only $V_{\pi + \delta}$ for the singlet and triplet channels
using the one-fold subtracted T-matrix equations as performed in
detail in Ref. \cite{fred99}, where we have used
\begin{eqnarray}
V^{(1)}_{\pi+\delta} (p',p\,;-\mu^2;k^2) =
V_\pi (p',p) + \lambda_{\mathcal{R}00} ~.
\label{v1pd}
\end{eqnarray}
It is worth to mention that, for $\mu \rightarrow \infty$, $V_\pi
(p',p) \rightarrow T_\pi(p',p\,;-\mu^2;k^2)$. In the present
framework, we are using directly as input
$T_\pi(p',p\,;-\mu^2;k^2)$ (see Eqs. (\ref{TPIVW3}) and
(\ref{vrenfinal})). The large values of $\mu$ in Fig. \ref{fig1}
arise due to the choice given by Eq. (\ref{v1pd}). We observe that
the large $\mu$ limit converges. For the singlet channel,
$V_{\pi+\delta}$ is not enough to describe the data. However, the
approach works reasonably well in the spin-triplet channel when we
consider Eq. (\ref{v1pd}) for the coupled channels with $\ell = 0$
and $\ell = 2$. Therefore, we will not consider the spin-triplet
channel further in the present work where we focus on the huge
discrepancy of the singlet channel. We leave a detailed analysis
of the coupled channels for a future work, where we also intend to
include the two-pion-exchange interaction.

\begin{figure}[t]
\includegraphics[width=10cm ,height=8cm]{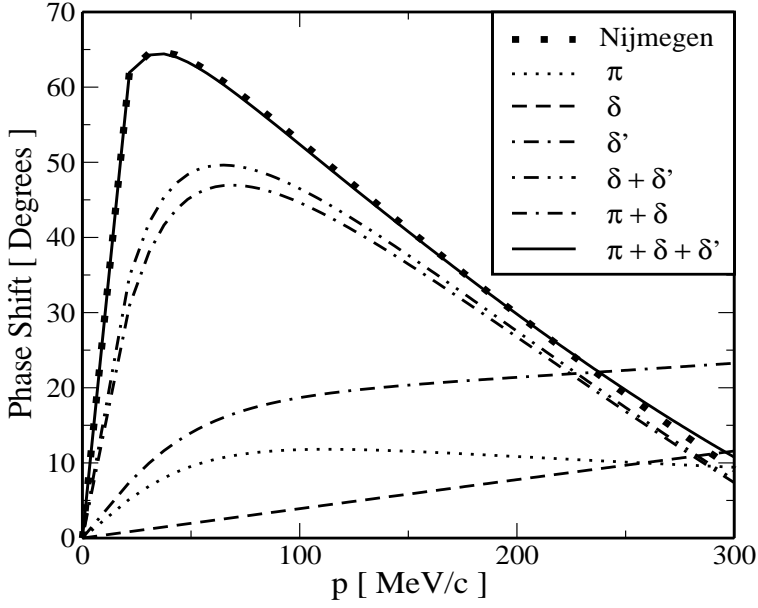}
\caption[dummy0] {Phase shifts for the singlet ($^1S_0$) for
different components of the effective interaction (\ref{vefts})
for $\mu=214$MeV/c. The solid line is the calculation with all
terms of the effective potential of (\ref{vrenfinal}).}
\label{fig2}
\end{figure}

In Figure~\ref{fig2} we present the results obtained when we
include $V_{\delta'}$. In order to evaluate the contributions of
the different terms in the effective potential for the parameters
of the best fit displayed in Table I, we calculate the $^1S_0$
phase shifts for the individual contributions of $V_\pi$,
$V_\delta$ and $V_{\delta'}$, as well as for the combinations
$V_{\delta+\delta'}$, $V_{\pi+\delta}$ and
$V_{\pi+\delta+\delta'}$. Below $p \sim 20$~MeV/c the different
calculations presents results that underestimate the data due to
the fact that $|a_s|<<$ 23.7~fm,
especially for $V_\pi$ and $V_{\pi+\delta}$, while for
$V_{\delta'}$ and $V_{\delta+\delta'}$ $a_s$ is better
approximated (see Table~\ref{tab1}). In the last two cases, as the
short-range part of the effective interaction is dominated by the
higher-order derivatives of the Dirac-delta, the calculation is
about enough to describe the phase shifts above $p\sim 150$~MeV/c.
Therefore, the most relevant contribution in this channel comes
from $V_{\delta+\delta'}$, while the  regular part of the pion
exchange potential appears at low energies as it provides the long
range part of the potential.

\begin{figure}[t]
\includegraphics[width=10cm ,height=8cm]{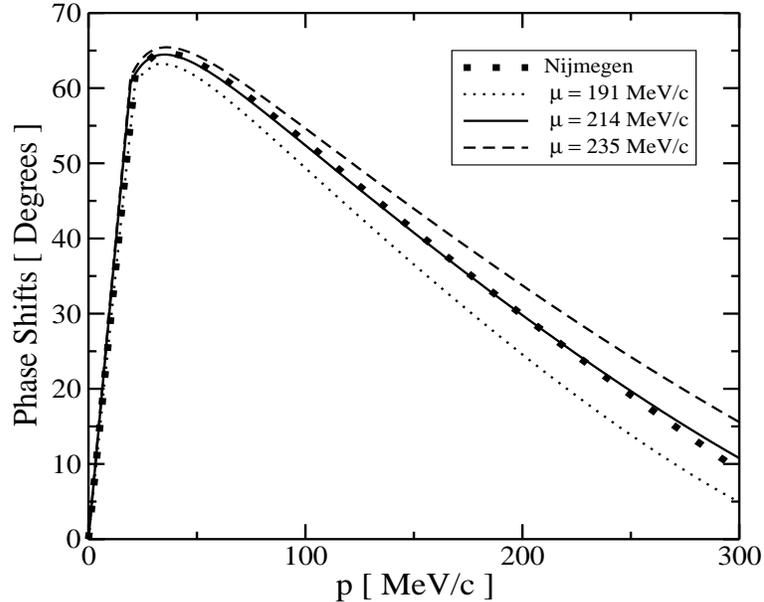}
\caption[dummy0] {The $\mu$ dependence of the singlet phase-shift
for $V_{\pi + \delta + \delta^\prime}$.} 
\label{fig3}
\end{figure}

 Although our approach is RG invariant, it is not simple to
explicitly incorporate RG invariance in the driving term of the
three-fold subtracted equations. In order to have an insight of
the effects included by the evolution of the driving term through
the CS equation, it is important to study how our fit depends on
$\mu$. For this purpose, we calculate the phase shifts for values
of $\mu$ around the best-fit value ($\mu = 214\rm MeV/c$) keeping
the $\lambda$'s intact, as shown in Fig. \ref{fig3}. We observe
that, at very low momenta the fit is not affected. For higher
momenta, the phase shifts present a monotonic behaviour: it
increases (decreases) as $\mu$ gets larger (smaller). The use of
the NRCS equation (\ref{ss3}) would cancel this effect in order to
keep the T-matrix, and thus the observables, invariant.

In conclusion, we show in  this letter how to apply the subtracted
scattering equations~\cite{plb00} to the NN singlet channel, when
more than one subtraction is required to renormalize the
corresponding interaction.

The approach consists of a regularization of the integrand of the
original scattering equation by using a given subtraction
procedure in the propagator at some energy scale with  where the
T-matrix is known. The subtraction scale can be moved without
modifying the calculated observables once the inhomogeneous term
of the subtracted scattering equation runs with the scale
according to the Non-Relativistic Callan-Symanzik (NRCS) equation.
In the present case, the boundary condition is determined by the
renormalized coupling constants and by the T-matrix of the OPE and
Dirac-delta potentials at the specified value of the subtraction
point. As a final remark, the subtracted scattering equation
method is recursive and can be easily extended when higher order
 derivatives of the Dirac-delta are present in the two-body
effective interaction.

{\bf Acknowledgements}
\\
This work was partially supported by Funda\c c\~ao de Amparo \`a
Pesquisa do Estado de S\~ao Paulo (FAPESP), Conselho Nacional de
Desenvolvimento Cient\'\i fico e Tecnol\'ogico (CNPq) and Funda\c
c\~ao de Amparo \`a Pesquisa do Estado do Rio de Janeiro (FAPERJ).
V.S.T thanks to FAEPEX/UNICAMP for financial support.

\end{document}